\documentclass[pre,preprintnumbers,amsmath,amssymb,twocolumn]{revtex4}
\usepackage{graphicx}
\usepackage{bm,color}
\usepackage{caption}
\usepackage{mathrsfs}
\usepackage{verbatim} 
\usepackage{tikz}

\definecolor{labelkey}{cmyk}{.4,.2,0,0}

\newcommand{\nn}{\nonumber}

\begin{document}

\title{On the role of electron-nucleus contact and microwave saturation in Thermal Mixing DNP}
\author{\bf  Sonia Colombo Serra$^1$, Alberto
Rosso$^{2}$ and Fabio Tedoldi$^1$}

\affiliation{\medskip
$^{1}$Centro Ricerche Bracco, Bracco Imaging Spa, via Ribes 5, 10010 Colleretto Giacosa (TO), Italy. \\
$^{2}$Universit\'e Paris-Sud, CNRS, LPTMS, UMR 8626, Orsay F-91405, France.\smallskip}
\begin{abstract}
We have explored the manifold physical scenario emerging from a model of Dynamic Nuclear Polarization (DNP) via thermal mixing under the hypothesis of highly effective electron-electron interaction. When the electron and nuclear reservoirs are also assumed to be in strong thermal contact and the microwave irradiation saturates the target electron transition, the enhancement of the nuclear polarization is expected to be considerably high even if the irradiation frequency is set far away from the centre of the ESR line (as already predicted by Borghini) and the typical polarization time is reduced on moving towards the boundaries of said line. More reasonable behaviours are obtained by reducing the level of microwave saturation or the contact between electrons and nuclei in presence of nuclear leakage. In both cases the function describing the dependency of the steady state nuclear polarization on the frequency of irradiation becomes sharper at the edges and the build up rate decreases on moving off-resonance. If qualitatively similar in terms of the effects produced on nuclear polarization, the degree of microwave saturation and of electron-nucleus contact  has a totally different impact on electron polarization, which is of course strongly correlated to the effectiveness of saturation and almost insensitive, at the steady state, to the magnitude of the interactions between the two spin reservoirs. The likelihood of the different scenario is discussed in the light of the experimental data currently available in literature, to point out which aspects are suitably accounted and which are not by the declinations of thermal mixing DNP considered here.
\end{abstract}

\maketitle
 
\section{Introduction}
\label{intro}

In the last decade Dynamic Nuclear Polarization (DNP) has established itself as a powerful technique to overcome the limited sensitivity of Nuclear Magnetic Resonance (NMR) \cite{PNAS JHAL}. More recently, as a consequence of the impressive experimental results over a wide area of applications, ranging from analytical \cite{SS1,SS2} to potentially diagnostic methods \cite{BA1, BA2, BA3}, the scientific community has started to deepen the existing theoretical knowledge about the physics of the polarization process \cite{Vega1, kock, Vega3}. Depending on the specific conditions of the experiment, the transfer of magnetic order from the electron to the nuclear system occurs by different mechanisms, named Solid Effect \cite{Abragam e Goldman, khut}, Cross Effect \cite{CE1, CE2, CE3, CE4} and Thermal Mixing \cite{Borghini PRL, Abragam e Goldman, Weck}.
The latter regime is believed to apply to those samples and experimental conditions typically exploited in biomedical applications \cite{JHAL2008}, which nowadays are attracting a large interest. 

The original theoretical description of low temperature DNP via Thermal Mixing (TM) is due to Borghini \cite{Borghini PRL} and re-proposed in a slighly different fashion by Abragam and Goldman in their famous review  \cite{Abragam e Goldman}.
The model is based on the hypothesis of \emph{(i)} very efficient spectral diffusion, \emph{(ii)} complete saturation of the irradiated Electron Spin Resonance (ESR) isocromate and \emph{(iii)} existence of a perfect contact between electrons and nuclei. 
This latter forcing the establishment of a common temperature between nuclear and electron reservoirs at any time. 

Despite the Borghini prediction qualitatively depicts some aspects of the experimental scenario, no information about the dynamics of the process is given, while the steady state nuclear polarization is always overestimated. The quantitative agreement is especially poor when moving from the centre to the edges of the ESR spectrum. In order to reduce the discrepancies between theory and experimental observations in the nuclear steady state behaviour, Jannin et al. \cite{JanninMW} has recently proposed a variant of the Borghini model where the irradiated ESR line portion is only partially saturated. Again, the dynamical problem has not been tackled.

A general methodology to compute the full time evolution of the nuclear polarization in the low temperature TM regime, relying on a mean field approach and based on a proper system of rate equations, has been described in  \cite{nostroPCCP}. 
In this work we exploited that mathematical treatment (briefly recalled in Section \ref{sectionModel}) for providing a comprehensive picture of the electron and nuclear polarization dynamics (including the relevant steady states), over the whole microwave spectrum, for different choices of the five time constants describing the basic interactions and relaxation mechanisms. In particular, under the assumption of an optimal electron spin-spin contact, the role of microwave power and electron-nucleus interaction was investigated. 

The numerical results presented in Section \ref{NumericalResults} point out how the hypothesis of partial saturation introdocued in \cite{JanninMW}, not only improves the agreement between TM theory an the experimental data of steady state nuclear polarization, but also predicts a more realistic behaviour for the frequency dependence of the nuclear build up time. A similar qualitative agreement is obtained by mantaining the full saturation assumption included in the original Borghini model and relaxing the constrain of perfect electron-nucleus contact, in presence of a weak, electron independent, nuclear spin lattice relaxation term. The two options considered, both fairly good in accounting for the behaviour of the nuclear reservoir, generate completely different scenario in respect of the electron polarization, as widely discussed  in Section \ref{Discussion} in the light also of the experimental observations available in literature.

In order to make the reading of the manuscript more fluent, three Appendixes collecting most of the relevant mathematics have been added at the end of the main text.

\section{Model overview}
\label{sectionModel}
A system of $N_n$ nuclear spin ${\bf I}$ ($I = 1/2$, Larmor frequency $\omega_n$) and $N_e$ electron spins ${\bf S}$ ($I = 1/2$, mean Larmor frequency $\omega_e$ ($\approx 10^3 \omega_n$) is considered. The electron frequency distribution (ESR line) is supposed to be  inhomogeneously broadened \footnote{The main contribution to the line broadening is assumed to be single ion anysotropy (\emph{i.e.} spreading of $g$-factors), while the influence of dipolar electron-electron interaction on the broadening is neglected. As far as a sample doped with 15 mM of trityl radical is concerned, the tipical ESR line width at T = 1.2 K and $B_0$ = 3.35 T is about 63 MHz, whereas the electron-electron dipolar interaction is about 2 MHz. Despite the electron dipolar interaction brings a negligible contribution to the line width, it plays an important role as source of spectral diffusion between different spin packets.} and can be conveniently decomposed in a sequence of $N_p$ narrow individual spin packets of frequency $\omega_i = \omega_e - \Delta_i$, width $\delta \omega$ and relative weight $f_i$ such that $\sum f_i = 1$ and $\sum f_i \Delta_i = 0$. The system is assumed to be ruled by five processes: microwave irradiation (with a characteristic time $T_{\text{1MW}}$), spectral difusion ($T_{2e}$), ISS process ($T_{\text{ISS}}$), electron spin-lattice relaxation ($T_{1e}$) and nuclear spin-lattice relaxation ($T_{1n}$) (see \cite{nostroPCCP} for detailed description). 

The rate $1/T_{\text{ISS}}$ describes, as an effective parameter, both the nuclear spin diffusion process and the interaction of two generic electrons belonging to packets $i$ and $i + \delta n_p$ (being $\delta n_p$ the number of packets corresponding to $\omega_n$) and a generic nucleus $n$. Electrons of the same packet are set identical by definition and characterized by a local polarization $P_{e, i}$, whereas a unique polarization $P_n$ and inverse temperature $\beta_n$ is assigned to the whole nuclear system:
\begin{equation}
P_n (t) = \tanh[\beta_n(t)\delta n_p)].
\label{tanheqPn}
\end{equation}

The system so defined is studied in the TM regime, where the spectral diffusion processes, mediated by the electron dipolar interaction, are far more efficient than any other process ($T_{2e} \rightarrow 0$). In this limit, even when the system is out of equilibrium because of the MW irradiation, a unique spin temperature is established at all times among the electron packets (Appendix \ref{appTM}). 
The electron polarization $P_{e, i}(t)$ can thus be written as:
\begin{equation}
P_{e, i}(t) = \tanh[\beta_e(t)(\Delta_i -c(t))],
\label{tanheq}
\end{equation}
where $c(t)$ and the unique inverse temperature $\beta_e(t)$ are time-dependent parameters.

The dynamics of $P_{e, i}(t)$ and consequently of $P_n(t)$ is determined not only by the highly efficient spectral diffusion but also by the remaining four processes. Their effect is described by the system of rate equations introduced in \cite{nostroPCCP} and here reported for convenience of the reader (under the assumption of finite rates for all the four processes).
\begin{eqnarray}
\label{rateeq}
 \frac{d P_{e, i}(t)}{dt}&=& \frac{P_0-P_{e, i}(t)}{T_{1e}} - \delta_{i, i_0} \frac{P_{e,0}(t)}{T_{1 \text{MW}}} \\
 &+& \frac{f_{i-\delta n_p} \Pi_- + f_{i+\delta n_p} \Pi_+ }{2 T_{\text{ISS}}}  \nonumber \\
 \frac{d P_n(t)}{dt}&=& \frac{P_{0n}-P_n(t)}{T_{1n}}- \frac{N_e}{2 T_{\text{ISS}} N_n} \sum f_i f_{i+\delta n_p} \Pi_n  \nonumber
\end{eqnarray}
where $\delta_{i, i_0}$ is a Kronecker delta and $\Pi_-=\Pi_-(i,t), \Pi_+=\Pi_+(i,t), \Pi_n=\Pi_n(i,t)$ are given by the expressions:
\begin{eqnarray}
\Pi_- &=& P_{e, i-\delta n_p}(t) \small{-}P_{e, i}(t)\small{-}P_n(t)\left[1\small{-}P_{e, i-\delta n_p}(t)P_{e, i}(t)\right] \nonumber \\ 
\Pi_+ &\small{=}& \Pi_n \small{=} P_{e, i+\delta n_p}(t) \small{-} P_{e, i}(t)\small{+}P_n(t)\left[1\small{-}P_{e, i+\delta n_p}(t)P_{e, i}(t)\right] \nonumber \\
\end{eqnarray}  

For numerical computation a discrete time step $dt$ is introduced:
\begin{equation}
dt = \frac{1}{W_{1\text{MW}} + W_{\text{ISS}} + W_e + W_n}.
\end{equation}
where $W_{1\text{MW}}$ = $N_e f_0/T_{1\text{MW}}$, $W_e$ = $N_e/T_{1e}$, $W_{\text{ISS}}$ = $N_e \sum f_i f_{i+\delta n_p}/T_{\text{ISS}}$ and $W_n$ = $N_n/T_{1n}$.
After each elementary evolution step according to Eq.(\ref{rateeq}), the effect of spectral diffusion (acting on a typical time scale $\delta t \approx T_{2e}\ll dt $) is accounted by imposing that the polarizations $P_{e, i}(t + \delta t)$ satisfy Eq.(\ref{tanheq}) and the conservation of the energy and total polarization: 
\begin{eqnarray}
\sum f_i \left[P_{e, i}(t\small{+}\delta t)\small{-}P_{e, i}(t)\right] = 0 \nonumber \\
\sum f_i \Delta_i \left[P_{e, i}(t\small{+}\delta t)\small{-}P_{e, i}(t)\right] = 0.
\end{eqnarray}

In this work we investigate three distinct regimes where, in addition to spectral diffusion, one or two more processes are assumed infinitely efficient.
\subsection{Regime I (`Borghini')} 
The evolution of the system is derived under the following assumptions:
\begin{itemize}
\item $T_{\text{ISS}} \rightarrow 0$: a perfect contact between the electron and the nuclear reservoirs which allows to establish a common electron-nucleus inverse temperature $\beta(t) = \beta_e(t) = \beta_n(t)$ at all times (see Appendix \ref{appTM});
\item $T_{1\text{MW}} \rightarrow 0$: a full saturation of the irradiated packet $i_0$ which corresponds to assume $P_{e, 0}(t) = 0$, so that $c(t) = \Delta_0$.
\end{itemize}
The steady state solution $P_{e, i}(t \rightarrow \infty)$, $P_n(t \rightarrow \infty)$  can be computed by solving numerically the known Borghini relation: 
\begin{equation}
\label{BorghiniRelation}
\sum f_i (\Delta_i - \Delta_0) P_{e, i}+ \Delta_0 P_0 - \omega_n \frac{N_n T_{1e}}{N_{e}T_{1n}}P_n = 0,
\end{equation}
which can be easily obtained by solving the system of equations describing the time evolution of the two energy reservioirs (Zeeman electron and non-Zeeman plus Zeeman nuclear contributions, reported in Equation C4 of \cite{nostroPCCP}) at the steady state, under the condition $c = \Delta_0$.

The dynamics of electron and nuclear polarizations can be obtained from the system of rate equations (\ref{rateeq}), conveniently adapted to this regime (Appendix \ref{appDynamicsI}). Moreover, it is possible to write the rate equation for the inverse temperature $\beta(t)$ (Appendix \ref{appDynamicsI}), whose solution is not an exponential function.     
\subsection{Regime II (`partial MW saturation')} 
The evolution of the system is derived under the following assumptions:
\begin{itemize}
\item $T_{\text{ISS}} \rightarrow 0$: a perfect contact between the electron and the nuclear reservoirs which imposes, as in regime I, $\beta(t) = \beta_e(t) = \beta_n(t)$; 
\item $T_{1\text{MW}} \neq 0$: an incomplete saturation of the irradiated packet $i_0$.
\end{itemize}
The steady state solution is now function of two variables $\beta$ and $c$ and can be evaluated by numerically solving the following system of two equations:
\begin{eqnarray}
\label{BorghiniRelationMW}
&&\sum f_i \frac{P_{0}-P_{e, i}}{T_{1e}} - f_0 \frac{P_{e, 0}}{T_{1 \text{MW}}} = 0 \nonumber \\ 
&&\sum \Delta_i f_i \frac{P_{e, i}}{T_{1e}}  + f_0 \Delta_0 \frac{P_{e, 0}}{T_{1 \text{MW}}} - \frac{N_n}{N_e} \hbar \omega_n \frac{P_n}{T_{1n}} = 0
\end{eqnarray} 
which is a generalized version of the Borghini relation, again obtained as steady state solution of the system of rate equations reported in Equation c4 of \cite{nostroPCCP}.

The evolution of $P_{e, i}(t)$ and $P_n(t)$ can be estimated by means of the system of rate equations (\ref{rateeq}) adapted for this regime (Appendix \ref{appDynamicsII}).
\subsection{Regime III (poor electron-nucleus contact)}
The evolution of the system is derived under the following assumptions:
\begin{itemize}
\item $T_{1\text{MW}} \rightarrow 0$: a full saturation of the irradiated packet $i_0$ which imposes $c = \Delta_0$:
\item $T_{\text{ISS}} \neq 0$: a poor contact between the electron and the nuclear reservoirs, modulated by the corresponding parameter $1/T_{\text{ISS}}$, which leads (in presence of leakage) to two different inverse temperatures for electrons and nuclei, \emph{i.e.} $\beta_e(t) \neq \beta_n(t)$.
\end{itemize}

The steady state solution is now function of two variables $\beta_e$ and $\beta_n$ and can be evaluated by numerically solving a system composed by the Borghini relation (Eq.(\ref{BorghiniRelation}), which holds also in this regime, but it is not sufficient to determine unambiguously $\beta_e$ and $\beta_n$) and the rate equation for $P_n(t)$. This latter, once imposing the stationary condition, writes:
\begin{equation}
\label{regimeSSIII}
P_n\small{=}\frac{\frac{N_e}{2 T_{\text{ISS}} N_n} \sum f_i f_{i+\delta n_p}\left(P_{e, i}\small{-}P_{e, i+\delta n_p}\right)+\frac{P_{0n}}{T_{1n}}}{\frac{N_e}{2 T_{\text{ISS}} N_n} \sum f_i f_{i+\delta n_p}\left(1-P_{e, i}P_{e, i+\delta n_p}\right)+\frac{1}{T_{1n}}} 
\end{equation}

The solution for $P_{e, i}(t)$ and $P_n(t)$ can be obtained from the system of rate equations (\ref{rateeq}), conveniently adapted for this regime (Appendix \ref{appDynamicsIII}). 

In the limit $T_{\text{ISS}}\gg T_{1e}$, the contact between the electrons and the lattice is more efficient than the contact between electrons and nuclei. The steady state polarization profile $P_{e, i}$ is then achieved in a typical time of the
order of $T_{1e}$ independentely from any feature of the  nuclear reservoir. As a consequence, the nuclear system `sees', through $T_{\text{ISS}}$, an electron thermal bath at constant temperature and the rate equation for $P_n(t)$ assumes
the linear form:
\begin{equation}
\label{PnAB}
P'_n (t) = A - B P_n (t) 
\end{equation}
where $A$ and $B$ are constant terms defined as:
\begin{eqnarray}
A = -\frac{N_e}{2 T_{\text{ISS}} N_n} \sum f_i f_{i+\delta n_p} \left(P_{e, i+\delta n_p}\small{-}P_{e, i}\right) + \frac{P_{0n}}{T_{1n}} \nn \\
B = \frac{N_e}{2 T_{\text{ISS}} N_n} \sum f_i f_{i+\delta n_p} \left(1\small{-}P_{e, i+\delta n_p}P_{e, i}\right) + \frac{1}{T_{1n}}\nonumber
\end{eqnarray} 
being $P_{e, i}$ given by the Borghini relation (Eq.(\ref{BorghiniRelation}) in absence of nuclei).
Its solution:
\begin{equation}
P_n(t) = \frac{A}{B}\left[1-\exp(-B t)\right].
\end{equation}
is an exponential function with a steady state $P_n = A/B$ and an exponential time constant equal to $1/B$.

\section{Numerical Results}
\label{NumericalResults}
The three regimes introduced in Section \ref{sectionModel} have been explored by computing a set of build up curves (\emph{i.e.} polarization \emph{versus} time) for different values of the significant parameters: $i_0$, $T_{1\text{MW}}$, $T_{\text{ISS}}$ and $T_{1n}$.
All the other parameters of the rate equations, when not differently stated, have been set as follows: $N_n/N_e$ = 1000, $T_{1e}$ = 1 s, $N_p$ = 15, $\delta n_p$ = 3 and $f_i$ defined according to a Gaussian function with a full width at half maximum $\Delta \omega_e$ = 63 MHz and truncated at 3$\sigma$. This set of parameters is choosen to represent a sample of [1-13C]-pyruvic acid doped with 15 mM trityl radical in a magnetic field $B_0$ = 3.35 T, at temperature $T$ = 1.2 K . Such well known mixture is an ideal prototype to be tested against the outcome of our calulations, since it was argued to polarize via TM \cite{JHAL2008} and has been studied experimentally in great detail \cite{JHAL2008, JHAL2010, JHALhighfield}.
The build up curves obtained from the numerical simulation have been fitted by the phenomenological law:
\begin{equation}
P(t)=P_0\left[1-\exp\left(-t/T_{\text{pol}}\right)^{\alpha} \right]
\end{equation}
where $P_0$ is the steady state value of the polarization, $T_{pol}$ is the polarization time and $\alpha$ is a stretching exponent. For $\alpha$ = 1, the usual exponential function is recovered.

\subsection{Regime I: Borghini model}
\label{regimeI}

\begin{figure*}
\includegraphics[width=17.2 cm]{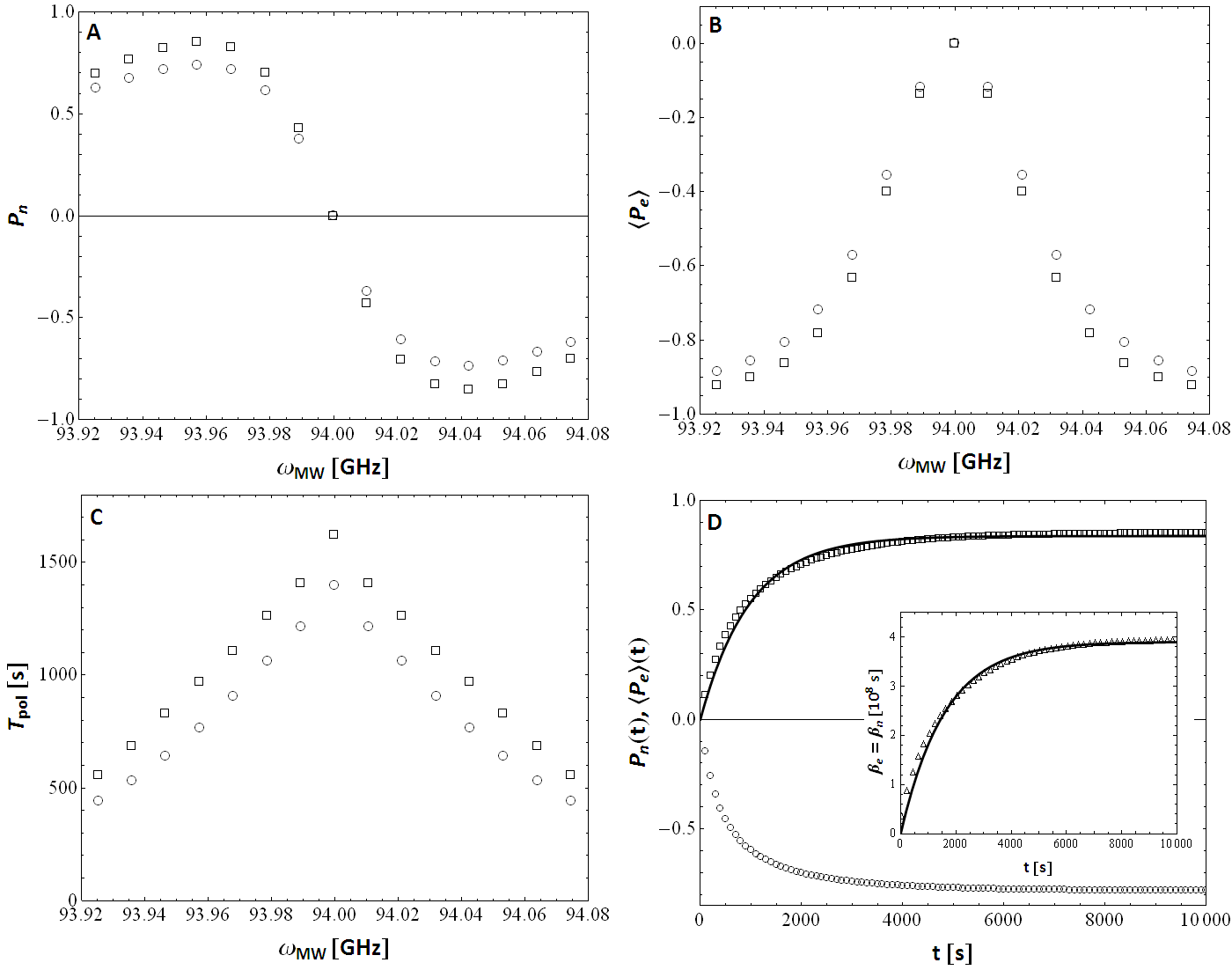}
\caption{Overview of nuclear and electron polarization in TM-DNP under the assumption defining regime I ($T_{2e}$ = 0 s, $T_{\text{ISS}}$ = 0 s, $T_{1\text{MW}}$ = 0 s) at 3.35 T and 1.2 K. The nuclei/electrons ratio has been set to $N_n/N_e = 1000$, whereas the electron longitudinal relaxation time is assumed to be $T_{1e}$ = 1 s. 
Panel A: Steady state nuclear polarization $P_n$ as a function of the irradiating frequency $\omega_{\text{MW}}$ in absence of leakage (squares) and with $T_{1n}$ = 10000 s (circles). Panel B: Average electron polarization $\langle P_e\rangle$ as a function of $\omega_{\text{MW}}$ in absence of leakage (squares) and with $T_{1n}$ = 10000 s (circles). Panel C: Nuclear polarization time $T_{\text{pol}}$ as a function of $\omega_{\text{MW}}$ in absence of leakage (squares) and with $T_{1n}$ = 10000 s (circles). Panel D: Nuclear and average electron polarization build up curves $P_n(t)$ (squares) and $\langle P_e(t)\rangle$ (circles) at $\omega_{\text{MW,opt}}$ in absence of leakage.  The growing curve of the inverse temperature $\beta(t)$ (triangles) is represented in the inset. 
The non linearity of the differential equations which regulate this regime is reflected in the lack of agreement between the calculated trend of $P_n(t)$ and $\beta(t)$ and the exponential best fittings (solid lines).}
\label{figureI}
\end{figure*}

The physical scenario emerging under the assumptions defining the regime I is summarized in \figurename~\ref{figureI}. Panel A shows the steady state nuclear polarization $P_n$ as a function of the microwave frequency $\omega_{\text{MW}}$ in absence of leakage and when $T_{1n}$ = 10000 s. The same curve is obtained by solving Eq.(\ref{BorghiniRelation}). The calculated values are generally higher than those experimentally observed especially when moving from the centre to the edges of the ESR line. A maximum nuclear polarization of 0.85 is reached in absence of leakage when the microwave frequency is set to $\omega_{\text{MW,opt}}$ = $\omega_e$ - 43 MHz (corresponding to the irradiation of the packet $i_0$ = 4). Leakage has only a moderate effect on the curve, leading to a 13 $\%$ reduction of the maximum polarization level for $T_{1n}$ = 10000 s.

Another interesting quantity for comparison with experiments (see Section \ref{Discussion}) is the average electron polarization $\langle P_e(t) \rangle$, defined as:
\begin{equation}
\langle P_e(t)\rangle=\sum f_i P_{e,i}(t).
\end{equation}
The steady state value $\langle P_e \rangle$ of this quantity is reported in panel B of \figurename~\ref{figureI}. When the irradiation frequency $\omega_{\text{MW}}$ is close to $\omega_e$, the ESR line is effectively saturated, \emph{i.e.} 
 $\langle P_e \rangle \approx 0$. Conversely when the irradiation frequency is set at the edges of the ESR line, $\langle P_e \rangle \rightarrow P_0$ because of the low weigth $f_i$ of the side packets. 
    
The dynamical evolution of the spin systems can be derived by means of Eq.(\ref{CLBorghini}) and Eq.(\ref{rateeqBorghini}). 
The behaviour of the nuclear polarization time as function of $\omega_{\text{MW}}$ is shown in \figurename~\ref{figureI},  panel C. The larger is the shift between $\omega_e$ and $\omega_{\text{MW}}$, the shorter is $T_{\text{pol}}$ or, in other words, the steady state is achieved faster when the edges of the ESR line are irradiated. 

Finally, on panel D of \figurename~\ref{figureI}  the nuclear and electron build up curves are displayed for $\omega_{\text{MW,opt}}$ (\emph{i.e.} with $i_0$ = 4) and no leakage starting from the thermal Boltzman equilibrium condition $P_n \approx$ 0 and $P_{e, i} = P_{0}$, $\forall i$. As introduced in the previous Section, the same inverse temperature $\beta(t)$ (see inset) characterizes both nuclear and electron reservoir. The dynamics is non exponential as can be noticed by the mismatch between the best fitting curve and the simulated data if a stretching exponential $\alpha$ = 1 is assumed. The function describing the dynamics of $\beta(t)$ is computed in Appendix \ref{appDynamicsI}. 
Despite nuclei and electrons share the same temperature over time, since the hyperbolic tangent is a non linear function, the nuclear and electron polarization build up times are slightly different, although both in the order of $10^3$ s. The initial condition $\beta (t=0) \approx 0$ originates as follows: 
\begin{itemize}
\item when MW are switched on the packet $i_0$ is immediately saturated ($P_{e, 0} = 0$) due to the assumption $T_{1\text{MW}} \rightarrow 0$;
\item the fast spectral diffusion ($T_{\text{2e}} \rightarrow 0$) imposes: $P_{e, i}(t=0) = \tanh \left[\beta(t=0)\left(\Delta_i-\Delta_0\right)\right]$;
\item the effective contact between electrons and nuclei gives: $\beta(t=0) = \beta_n(t=0) \approx 0$.
\end{itemize}

\subsection{Regime II: partial MW saturation}
\label{regimeII}

\begin{figure*}
 \includegraphics[width=17.2 cm]{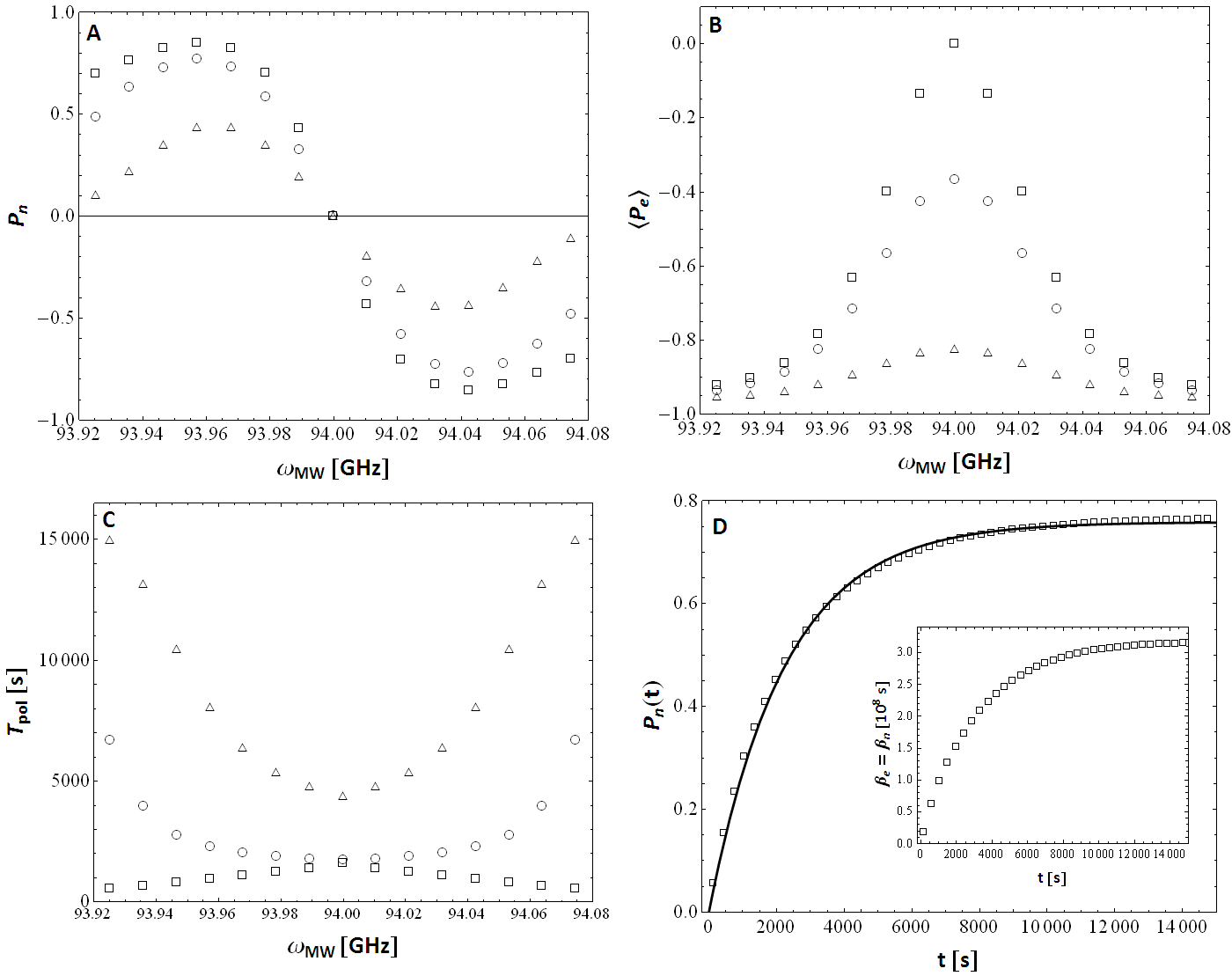}
\caption{Overview of nuclear and electron polarization in TM-DNP under the assumption defining regime II ($T_{2e}$ = 0 s, $T_{\text{ISS}}$ = 0 s, $T_{1\text{MW}} \neq 0$) at 3.35 T and 1.2 K, with $N_n/N_e = 1000$, $T_{1e}$ = 1 s and in absence of leakage. Panel A: Steady state nuclear polarization $P_n$ as a function of the irradiating frequency $\omega_{\text{MW}}$ with $T_{1\text{MW}}$ = 0 (squares), 0.1 s (circles) and 1 s (triangles). Panel B: Average electron polarization $\langle P_e\rangle$ as a function of $\omega_{\text{MW}}$ with $T_{1\text{MW}}$ = 0 (squares), 0.1 s (circles) and 1 s (triangles). Panel C: Nuclear polarization time $T_{\text{pol}}$ as a function of $\omega_{\text{MW}}$ with $T_{1\text{MW}}$ = 0 (squares), 0.1 s (circles) and 1 s (triangles). Panel D: Nuclear polarization build up curve $P_n(t)$ for $T_{1\text{MW}}$ = 0.1 s at $\omega_{\text{MW,opt}}$. The mismatch between numerical data and exponential best fitting (solid line) points out the non linearity of the phenomenon. The inset shows the corresponding  growth of the inverse temperature $\beta(t)$.}
\label{figureII}
\end{figure*}

An overview of the regime characterized by a partial saturation of the ESR line is presented in \figurename~\ref{figureII}. In panel A and B, $P_n$ and $\langle P_e\rangle$ as function of $\omega_{\text{MW}}$ are shown for $T_{1\text{MW}}$ = 0, 0.1 and 1 s, \emph{i.e.} moving from high to low MW power. The effect of a partial saturation is twofold: on one side a significant reduction of the maximum nuclear and electron polarization values is observed. On the other side a clipping of the wings of the steady state nuclear polarization curve occurs, making this latter more similar to the DNP spectrum experimentally observed in the prototype trityl doped sample studied in \cite{JHAL2008}. That uncomplete MW saturation can be invoked to better account for nuclear steady state data was pointed out previously in \cite{JanninMW}. Thank to this assumption the authors succeeded in fitting the DNP spectrum of a [1-13C]-sodium acetate sample doped with TEMPO, a free radical characterized by a much shorter $T_{1e}$ with respect to trityls and by a higher anisotropy of the $g$-tensor, resulting in turn in a wider ESR spectrum ($\approx$ 200 MHz \emph{vs} 60 MHz of trityls).

The behaviour of the polarization time versus $\omega_{\text{MW}}$ (panel C) is, interestingly, completely different from  regime I. 
As long as the irradiation frequency is close to $\omega_e$, $T_{\text{pol}}$ is relatively short, becoming longer and longer on moving towards 
the edges of the ESR line (from 1800 to 6800 s when $T_{1\text{MW}}$ = 0.1 s and from 4800 to 15000 s when $T_{1\text{MW}}$ = 1 s). As expected, longer is $T_{1\text{MW}}$, less effective the polarization mechanism is.

The time evolution of $P_n$  represented in panel D, as well as the build up curve of $\beta$ (inset) are similar to those obtained in regime I, with a non exponential behaviour (a rigorous demonstration is not reported in this case) and a typical time constant in the order of $10^3$ s. The build up of the electron polarization instead is somehow more complex and characterized by  two different time scales. The detail of such behaviour are analyzed in Appendix \ref{kineticPe}. 

\subsection{Regime III: poor electron-nucleus contact}
\label{regimeIII}
\begin{figure*}
 \includegraphics[width=17.2 cm]{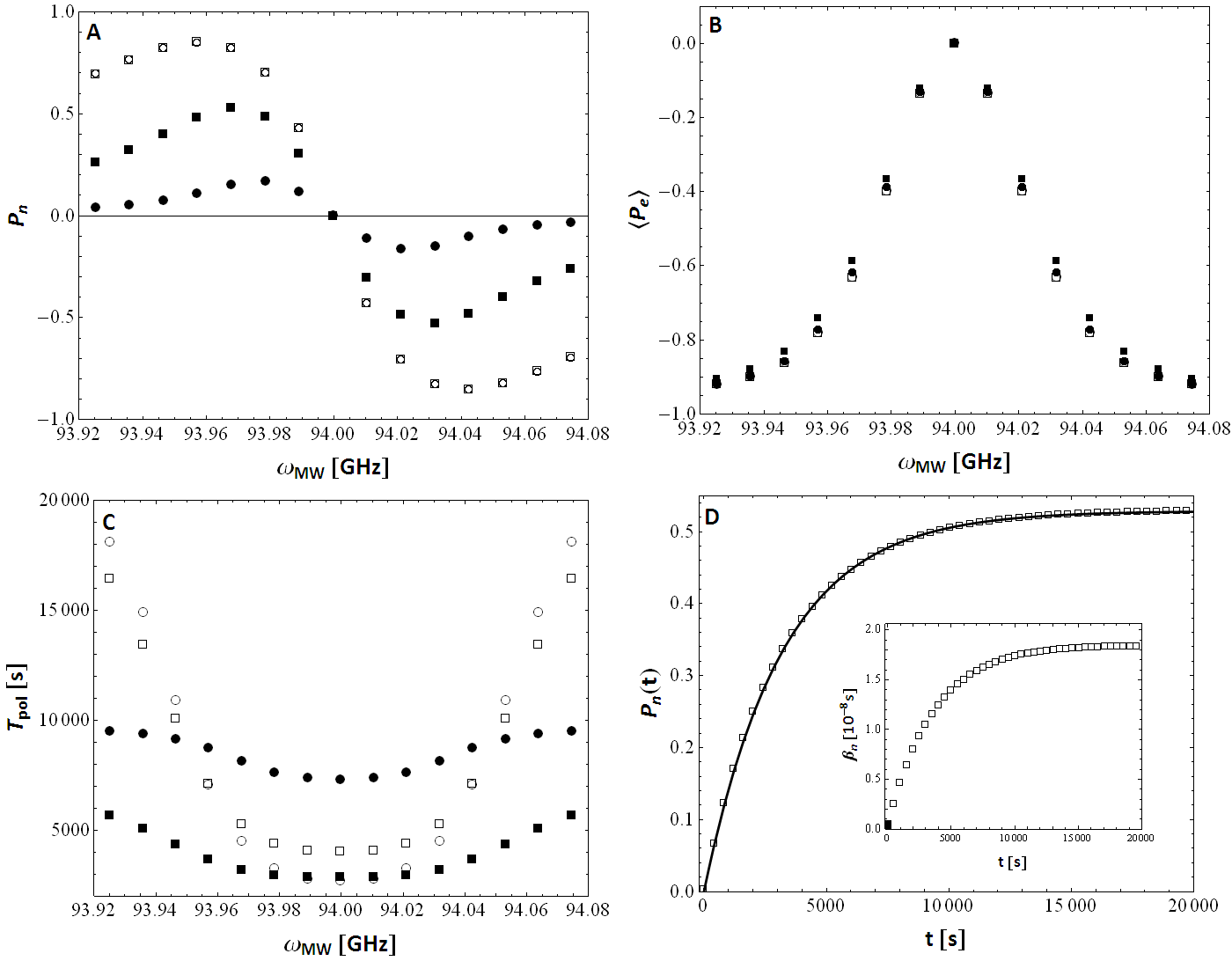}
\caption{Overview of nuclear and electron polarization in TM-DNP under the assumption defining regime III  ($T_{2e}$ = 0 s, $T_{1\text{MW}}$ = 0 s, $T_{\text{ISS}} \neq$ 0 s) at 3.35 T and 1.2 K, with $N_n/N_e = 1000$, $T_{1e}$ = 1 s. Panel A: Steady state nuclear polarization $P_n$ as a function of the irradiating frequency $\omega_{\text{MW}}$ with $T_{\text{ISS}}$ = 0.1 s (squares) or 1 s (circles) and in absence of leakage (empty symbols) or with $T_{1n}$ = 10000 s (filled symbols). Panel B: Average electron polarization $\langle P_e\rangle$ as a function of $\omega_{\text{MW}}$ with $T_{\text{ISS}}$ = 0.1 s (squares) or 1 s (circles) and in absence of leakage (empty symbols) or with $T_{1n}$ = 10000 s (filled symbols). Panel C: Nuclear polarization time $T_{\text{pol}}$ as a function of $\omega_{\text{MW}}$ with $T_{\text{ISS}}$ = 0.1 s (squares) or 1 s (circles) and in absence of leakage (empty symbols) or with $T_{1n}$ = 10000 s (filled symbols). Values represented with empty circles are scaled of a factor of $\frac{1}{10}$. Panel D: Nuclear polarization build up curves $P_n(t)$ (squares) at $\omega_{\text{MW,opt}}$, $T_{\text{ISS}}$ = 0.1 s and  $T_{1n}$ = 10000 s. The solid line represents the best fit to an exponential function, while the inset displays the corresponding growth of the inverse temperature $\beta_n(t)$.}
\label{figureIII}
\end{figure*}

The third regime is characterized by a finite contact rate between nuclei and electrons. Nuclear and electron polarization were computed for $T_{\text{ISS}}$ = 0.1 and 1 s both in absence of leakage and with $T_{1n}$ = 10000 s. The main results are shown in \figurename~\ref{figureIII}, following the same scheme of the regimes discussed above. 
Panel A shows the dependence of the nuclear polarization on $\omega_{\text{MW}}$. In absence of leakage, as already discussed in \cite{nostroPCCP}, the contact rate $T_{\text{ISS}}$ does not affect the steady state but only the dynamics and thus the two curves at different $T_{\text{ISS}}$  overlap. In presence of leakage $P_n$ is reduced. The longer is $T_{\text{ISS}}$ the higher the decrease is. Moreover, the reduction is more significant at the edges of the DNP spectrum, a behaviour that becomes clear in the light of panel C, where $T_{\text{pol}}$ is shown to be strongly increased at the wings of the spectrum. As a consequence, nuclear relaxation (with rate $T_{1n}$ = 10000 s) becomes a strong competing mechanism with respect to the ISS process, forcing $P_{n}$ towards a lower steady state value. 

Panel B highlights a rather interesting feature of regime III: the steady state electron polarization is almost unaffected either by $T_{\text{ISS}}$ and $T_{1n}$. This indicates that the nuclear system, for sufficiently high values of  $T_{\text{ISS}}$, is only a spectator of the electrons re-arrangement under MW irradiation, playing no active roles in the evolution of the electron systems towards its equilibrium. Evolution that proceeds through a two-step process is discussed in Appendix \ref{kineticPe}.
Nuclei have a `delayed response' characterized by a time constant in the order of  $N_n/(N_e T_{\text{ISS}})$ (about $10^4$ - $10^5$ s for the set of parameters used here) and by an exponential shape as confirmed by the good match between the fitting and the simulated data in panel D and as demonstrated in Section \ref{sectionModel}. The exponential time course of $P_n(t)$ stems from the linear rate equation (\ref{PnAB}), that further remarks the passive role of the nuclear reservoir in the polarization process of electrons. Correspondingly the nuclear inverse temperaure $\beta_{n}(t)$ builds up (inset of panel D) towards a steady state value that, in presence of  leakage, is substantially different from the end value of $\beta_e(t)$. One has, for instance,  $\beta_{n}= 1.84 $ x $ 10^8 $ s \emph{vs} $\beta_e(t)=3 $ x $ 10^8 $ s for $T_{\text{ISS}}=0.1$ s and $T_{1n}=10000$ s.

\section{Discussion and Conclusions}
\label{Discussion}

\begin{figure*}[htbp]
 \includegraphics[width=18 cm]{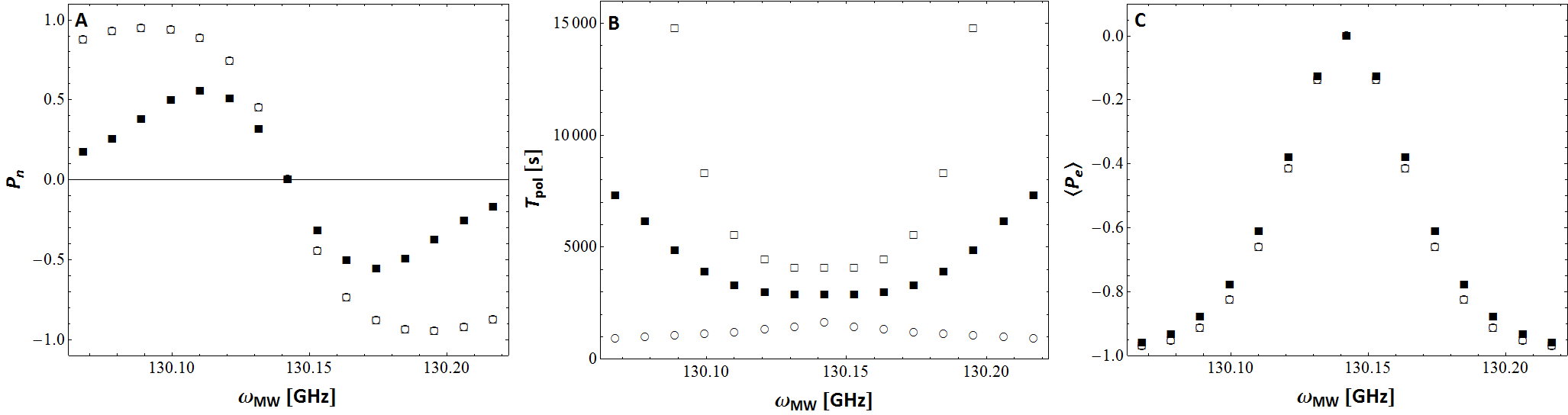}
\caption{Steady state nuclear polarization (panel A), nuclear polarizatione time (panel B) and steady state electron polarization (panel C) as a function of the microwave frequency at $B_0$ = 4.64 T and $T$ = 1.2 K in the Borghini regime (empty circle) and in the finite electron nucleus contact regime ($T_{\text{ISS}}$ = 0.1 s, without leakage (empty squares) and with $T_{1n}$ = 10000 s (filled squares)). Remaining parameters are set as follows $\delta \omega_e$ = 63 MHz, $T_{1e}$ = 1 s, $T_{1n} = \infty$, $N_n/N_e$ = 1000, $\delta n_p$ = 3, $N_p$ = 15.}
\label{HighField}
\end{figure*}
The original description of the TM mechanism proposed by Borghini and here analyzed in depth in terms of electron and nuclear polarization, polarization times and dynamics of the inverse spin temperatures has only a partial qualitative overlap with the experimental observations reported in literature. As, for example, pointed out in \figurename~8 of reference \cite{JHAL2008}, the Borghini model overestimates the final values of $P_n$ especially at the edges of the ESR line, leading to an unsatisfactory shape of the DNP spectrum ($P_n$ \emph{versus} $\omega_{\text{MW}}$). Similarly, our computation of the model, even when the MW frequency is 3 times $\sigma$ lower than $\omega_e$  (approx 80 MHz with our choice of parameters) and consequently the electron population of the corresponding energy levels is very low, predicts a very high enhancement of the nuclear polarization which is quite unrealistic and - more important - it is not experimentally observed. Conversely, by relaxing either the constraint of a complete MW saturation or the constraint of a perfect electron-nucleus contact, lower $P_n$ values and sharper DNP spectrum are obtained (panels A in \figurename~\ref{figureII}, \ref{figureIII}).

Furthermore, in the Borghini regime, the dependence of the efficiency of the polarization transfer on the microwave frequency ($T_{\text{pol}}$ \emph{vs} $\omega_{\text{MW}}$, panel C in \figurename~\ref{figureI}) disagrees with the experimental observations reported for a sample of [1-13C]-pyruvic acid doped with 10 mM of trityl. In Figure 5 of ref. \cite{JHAL2010} in fact Macholl et al. showed that $T_{\text{pol}}$ is relatively short as long as $\omega_{\text{MW}}$ is set between the two values corresponding to the positive and negative maximum of nuclear polarization (DNP optimum frequencies) whilst becoming longer and longer on moving  towards the edges of the ESR line. Remarkably, the correct qualitative behaviour of the polarization time is recovered under the assuptions underlying both regime II and regime III, as shown in panels B of \figurename~\ref{figureII} and \figurename~\ref{figureIII}. This type of dependence of $T_{\text{pol}}$  on  $\omega_{\text{MW}}$ looks not restricted to the reference sample and magnetic field value considered so far, but rather general for DNP experiments performed at very low temperature. Similar behaviours have been reported in fact for a sample of [1-13C]-pyruvic acid doped with trityl 18.5 mM at 1.2 K and 4.64 T, corresponding to an electron Larmor frequency of 130 GHz (Figure 3 in reference \cite{JHALhighfield}), as well as  for [1-13C]-labelled acetate doped  with TEMPO  50 mM at both 3.35 T (Figure 1 of \cite{Jannin2008}) and at 5 T (corresponding to $\omega_e \approx$ 140 GHz, Figure 2 of \cite{Jannin2008}) at a temperature of 1.2 K. The robustnees of the predictions of the model analyzed here against magnetic field strength has been verified for both regime II (data not shown) and regime III (\figurename~\ref{HighField}) by repeating the computation for our reference sample at higher field (parameters were set as follow, according to reference \cite{JHALhighfield}: $B_0$ = 4.64 T, $T$ = 1.2 K, $T_{1e}$ = 1 s, $\Delta \omega_e$ = 63 MHz). The extension of our calculations to a model representing different radicals and eventually higher temperatures, for comparison with the experimental observations achieved under such conditions \cite{Jannin2008, Griffin1, Griffin2, Griffin3}, will be faced in a next dedicated study.

Up to here, as long as the nuclear parameters only ($P_n$ and $T_{\text{pol}}$) are considered, regime II and III are both in qualitative agreement with the experimental observations and nearly superimposable \cite{Jannin2008, JHAL2010, JHALhighfield}. 
Actually the two regimes are very different, as can be understood from panels B and D of \figurename~\ref{figureII} and \figurename~\ref{figureIII}. For limited MW power and $T_{\text{ISS}}$ = 0, the nuclear system
and the electron one share always the same inverse temperature, generally lower than the achieved $\beta$ in case of full saturation. In fact, as shown in panel B of \figurename~\ref{figureII}, $\langle P_{e} \rangle$ 
tends to the frequency-independent equilibrium value $P_0$ when $T_{1\text{MW}}$ increases, as the competition between the MW pumping and the electron spin-lattice relaxation unbalances the steady state towards the Boltzman equilibrium. 
Being the electron system weakly affected by MW irradiation, it is not anymore a forceful source of polarization for nuclei.

On the other hand, in case of finite electron-nucleus contact and $T_{1\text{MW}}=$ 0, the electron system under the effect of the saturating MW pumping reaches in a short time a $quasi$-stationary polarization profile, characterized by an inverse temperature $\beta_e$ that slowly evolves while cooling the nuclear reservoir. In absence of leakage the nuclear system sees only the pre-thermalized electron reservoir and, with a characteristic time dependent on the contact ratio $1/T_{\text{ISS}}$ reaches a final inverse temperature $\beta_n = \beta_e$. In  presence of leakage the nuclear reservoir is on one side in thermal exchange with the electron system at $\beta_e$ and, on the other side, with the lattice at $\beta_L \propto 1/T$. The final nuclear inverse temperature $\beta_n$ is a trade-off value between $\beta_e$ and $\beta_L$ and, as well as nuclear build up time, depends on the two contact parameters $T_{\text{ISS}}$ and $T_{1n}$. 

In order to discriminate which scenario fits better with the experimental observations, data about the behaviour of electrons must be considered. A valid attempt to characterize the electron system was made by Ardenkjaer-Larsen and collaborators and it is reported in \cite{JHAL2008, JHALhighfield}. By measuring the shift of the $^{13}$C resonance line ($M_1$) caused mainly by the dipolar fields associated to the polarized paramagnetic centres, the authors indirectly estimated the average electron polarization $\langle P_e \rangle$ according to \cite{Abragam e Goldman}:
\begin{equation}
M_1=\frac{2}{3}\pi \xi \gamma_e \gamma_n \hbar N_e \langle P_e \rangle
\end{equation}
where $\xi$ is a coefficient which depends on the shape of the sample, $\gamma_e$ is the electron gyromagnetic ratio, $\gamma_n$ is the nuclear gyromagnetic ratio and $N_e$ is the number of electrons per unit volume. 

\begin{figure*}[htbp]
 \includegraphics[width=13.8 cm]{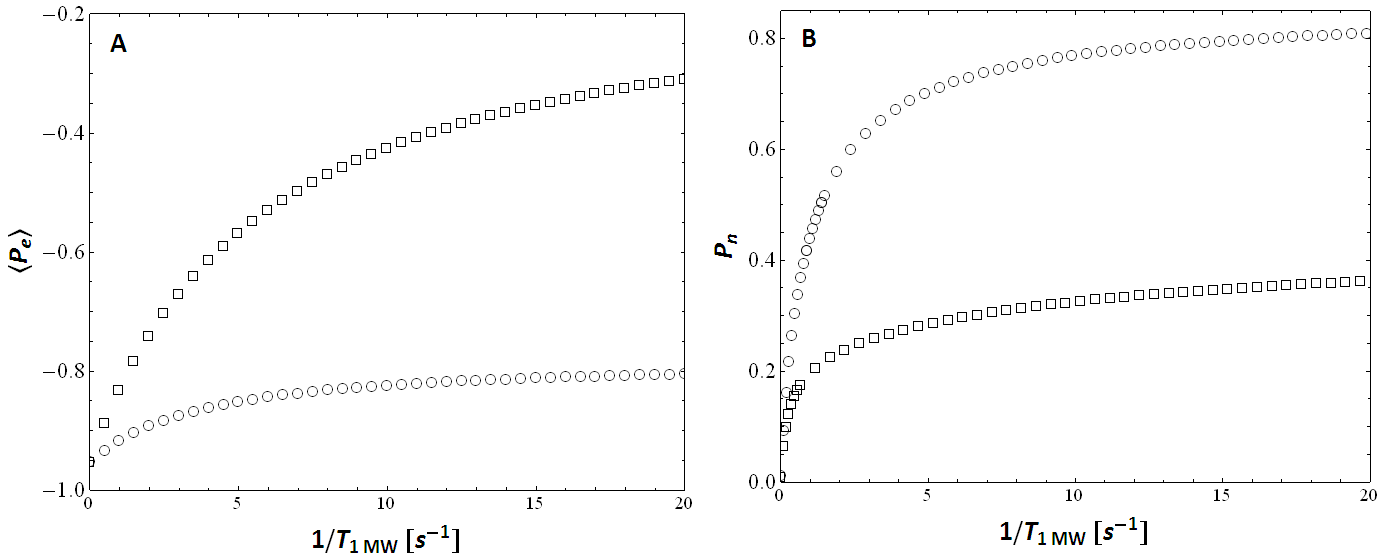}
\caption{Steady state polarization of the electron (panel A) and nuclear (panel B) spin systems as a function of the microwave power, expressed by the parameter $T_{1\text{MW}}$, for $\omega_{\text{MW}}$ = $\omega_{\text{MW,opt}}$ (circles) and $\omega_{\text{MW}} = \omega_e - \delta \omega$ (squares). Remaining parameters are set as follows: $T_{2e}$ = 0 s, $T_{\text{ISS}} = 0$, $T_{1e}$ = 1 s, $T_{1n} = \infty$, $N_n/N_e$ = 1000, $\delta n_p$ = 3, $N_p$ = 15.}
\label{figureMW}
\end{figure*}


In particular, in \cite{JHALhighfield} (Figure 4-6) the dependence of the nuclear shift, and thus indirectly of $\langle P_e \rangle$, on the MW frequency and power was measured \footnote{The experiments in \cite{JHALhighfield} have been performed at 4.64 T while the results of our calculation presented in Section \ref{NumericalResults} have been obtained at 3.35 T. We have however repeated the calculation of $\langle P_e \rangle$ at 4.64 T and found that its qualitative behaviour is not affcted by the intensity of $B_0$, as shown in \figurename~\ref{HighField} panel C.}.

The average electron polarization $\langle P_e \rangle$ at a fixed MW power (Figure 5 and 6 of \cite{JHALhighfield}) was found to depend on $\omega_{\text{MW}}$ with a behaviour similar to that reported in panel B of \figuresname~\ref{figureI}, \ref{figureII}, \ref{figureIII}, where the degree of electron saturation is higher for $\omega_{\text{MW}} = \omega_e -\delta \omega$ and lower when moving towards the edges of the ESR line and, as expected, when the MW power is reduced. 
It is worth to notice that $M_1$ increases rapidly at low microwave power and then reaches a plateau for power on the order of 40 - 60 mW. For a direct comparison of experimental (Figure 4 of \cite{JHALhighfield}) and computational data, the dependence of simulated levels of $P_n$ and $\langle P_e\rangle$ as function of the MW power, expressed by $T_{1\text{MW}}$, for different values of $\omega_{\text{MW}}$ is reported in \figurename~\ref{figureMW}. The same qualitative behaviour is obtained in experimental and calculated data. In numerical simulations the plateau is reached for $T_{1\text{MW}} \leq$ 0.05 - 0.1 s, whereas experimentally a plateau of $M_1$ is reached above few tens of mW. Such values are lower than the power level commonly used in DNP experiments at low temperature ($T \approx$ 1.2 K), thus suggesting that the assumption of full saturation is more appropriate than the hypothesis of partial saturation in interpreting DNP results collected on trityl doped samples in this temperature range. In \cite{JanninMW}, Jannin et. al. argued that the increase of the microwave power could lead, in the TEMPO doped sample considered, to a heating of the thermal bath which competes with the polarizing action of the MW themselves, affecting the equilibrium nuclear polarization and ending up in lower steady state $P_n$ values. 
Such an argument can not be extended to explain the observations on trityl doped samples, as the heating effect would also affect the equlibrium electron polarization, contributing positively to the total saturation of the ESR line. In that case, $\langle P_e \rangle$ should go to zero on increasing the MW power (1/$\beta_e \rightarrow \infty$), instead of going to the low temperature plateau observed in \cite{JHALhighfield}. Thus, the assumption of partial MW saturation, although successful in improving the description of DNP from the nuclear point of view, shows an intrinsic weakness in accounting for the electron behaviour of trityl doped samples (different conclusions may apply to DNP samples doped with different radicals, such as TEMPO, provided that $\langle P_e \rangle \rightarrow$ 0 on increasing the irradiation power). The model of finite electron-nucleus contact on the other hand, has similar capability in describing the nuclear system, but without explicitly contradicting the experimental behaviour of electrons . 
Overall, given the low temperature DNP experimental data available so far on the target compound considered here, relaxing the condition $T_{\text{ISS}}$ = 0 appears more promising than removing the saturation condition $T_{1\text{MW}}$ = 0.

\begin{figure*}[htbp]
 \includegraphics[width=13.8 cm]{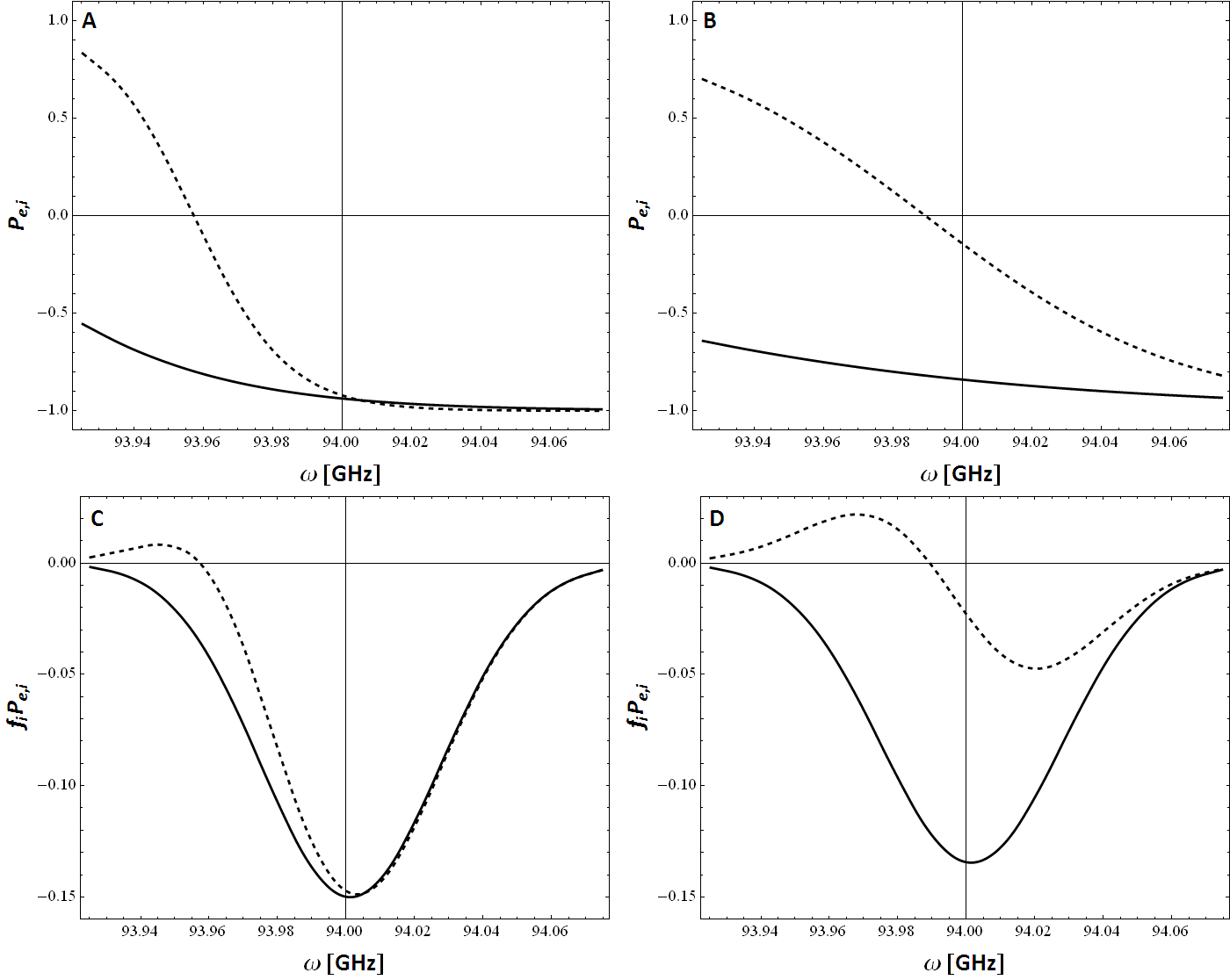}
\caption{Steady state electron polarization $P_{e,i}(t\rightarrow\infty)$ as function of the electron frequency $\omega$ in regime of partial saturation ($T_{1\text{MW}}$ = 1 s, solid line) and of non perfect electron-nucleus contact ($T_{\text{ISS}}$ = 1 s, dashed line)  for $\omega_{\text{MW}}$ = $\omega_{\text{MW,opt}}$ (panel A) and for $\omega_{\text{MW}} = \omega_e - \delta \omega$ (panel B). Weighted electron polarization $f_i P_{e,i}(t\rightarrow\infty)$ as function of the electron frequency $\omega$  in regime of partial saturation ($T_{1\text{MW}}$ = 1 s, solid line) and of non perfect electron-nucleus contact ($T_{\text{ISS}}$ = 1 s, dashed line)  for $\omega_{\text{MW}}$ = $\omega_{\text{MW,opt}}$ (panel C) and for $\omega_{\text{MW}} = \omega_e - \delta \omega$ (panel D). Remaining parameters are set as follows: $T_{2e}$ = 0 s, $T_{1e}$ = 1 s, $T_{1n} = \infty$, $N_n/N_e$ = 1000, $\delta n_p$ = 3, $N_p$ = 15.}
\label{figureMW2}
\end{figure*}

An elegant experimental test for better judging the physical meaningfulness of regime II and III would consist in measuring the electron polarization profiles $f_i P_{e,i}(\omega_{\text{MW}})$. The expected trends for the two regimes are shown in \figurename~\ref{figureMW2} for $\omega_{\text{MW}}$ = $\omega_{\text{MW,opt}}$ (panel A and C) and for $\omega_{\text{MW}} = \omega_e - \delta \omega$ (panel B and D) and described by Eq.(\ref{tanheq}): under the assumption of regime II $\beta_e = \beta_n$ and no packets are fully saturated, whereas in regime III in absence of leakage $\beta_e \neq \beta_n$ and the irradiated packet is characterized by $P_{e, 0} = 0$. Especially when $\omega_{\text{MW}}$ is set close to $\omega_e$ the electron profile of the two regimes are considerably different. 

In summary we have presented the articulated picture of thermal mixing DNP generated by the five parameters model introduced in \cite{nostroPCCP}, in the limit where $T_{2e}$ = 0.
Three cases in particular have been discussed in detail: the Borghini regime, characterized by a strong saturation of the ESR line and by a perfect contact between electrons and nuclei ($T_{1\text{MW}}$ and $T_{\text{ISS}}$ = 0), the regime of partial saturation of the ESR line ($T_{1\text{MW}} \neq$ 0 and $T_{\text{ISS}}$ = 0) and the regime of finite electron-nucleus contact ($T_{1\text{MW}}$ = 0 and $T_{\text{ISS}} \neq$ 0).
The former regime has been shown to be less accurate in accounting for the available experimental observations, whereas the latter two are both capable of properly capturing more features of the nuclear spin dynamics, whilst predicting different behaviour for the electron system. Additional dedicated experiments would be desirable in order to clarify which of the two predictions gives a better picture of the physical reality, although the finite electron-nucleus contact regime looks more consistent than partial saturation in describing the behaviour of trityl doped samples on varying the MW irradiation power. 

The theoretical picture exposed in this work cannot capture by definition those polarization phenomena driven by the Solid Effect or by the Cross Effect. Moreover it is still unable to describe some facts observed in experiments where the thermal mixing mechanism is expected to dominate, such as the inverse dependence of the nuclear steady state polarization on electron concentration, observed systematically when the ratio $N_e/N_n$ exceeds a certain value. The statistical approach introduced in \cite{nostroPCCP} however, can be extended to explore regimes with limited efficiency of the electron-electron interaction or, in other words, with limited thermal contact between different electronic packets. Moreover, the approach is flexible enough to allow the introduction of additional interaction terms. By exploiting these residual opportunities, we are confident that also the still unexplained behaviours will find suitable interpretation within the general framework of thermal mixing.

\section{Acknowledgement}
\label{Acknowledgement}
This study has been supported in part by Regione Piemonte (POR FESR 2007/2013, line I.1.1), by the COST Action TD1103 (European Network for Hyperpolarization Physics and Methodology in NMR and MRI) and by ANR grant 09-BLAN-0097-02.

\appendix

\section{Spin temperature in the Thermal Mixing regime}
\label{appTM}

Abragam and Goldman \cite{Abragam e Goldman} gave a description of TM DNP based on the separation between electron Zeeman and non-Zeeman contributions in the magnetic Hamiltonian of the system.

Starting from such hamiltonian one may derive the energy of a single electron spin $S_i$ (belonging to packet $i$) associated to its two possible states: \emph{up} ($\uparrow$) of energy $E^{\uparrow}_i =  \hbar/2 (\omega_e - \Delta_i)$ and \emph{down} ($\downarrow$), $E^{\downarrow}_i = - \hbar/2 (\omega_e - \Delta_i$). When MW are off the system is at thermal equilibrium with the lattice, at an inverse temperature $\beta_L = \hbar/(2 k_B T)$ (where $k_B$ is the Boltzman constant) and the probability for the spin $S^i$ to be in the state \emph{up} is given by the Boltzman weight:
\begin{equation}
p_i^{\uparrow} \propto \exp\left[-\beta_L(\omega_e - \Delta_i )\right]. \nonumber
\end{equation}
When MW are on, the system is out of equilibrium. If now the existence of a unique temperature among the different packets is postulated as in  \cite{Abragam e Goldman}, the  probability $p_i^{\uparrow}$ can be expressed in terms of a generalized Boltzman weight:  

\begin{equation}
p_i^{\uparrow} \propto \exp\left[-\left(\frac{\hbar \omega_e}{2 k_B T_{\alpha}}-\frac{\hbar \Delta_i}{2 k_B T_{\beta}}\right)\right] = \exp\left[-(\alpha \omega_e - \beta \Delta_i )\right] \nonumber
\end{equation}
where the two parameters $\alpha = \hbar/(2 k_B T_{\alpha})$ and $\beta=\hbar/(2 k_B T_{\beta})$ are normally referred as Zeeman and non Zeeman inverse temperature respectively.
The polarization of the spin $S^i$ can be then written as:
\begin{equation}
P_i = \frac{p^{\uparrow}_i-p^{\downarrow}_i}{p^{\uparrow}_i+p^{\downarrow}_i} = -\tanh\left[\alpha \, \omega_e - \beta \, \Delta_i\right]. \nonumber
\end{equation}   

The same espression can be derived by observing that, whenever a process much faster than the other events ruling the system exists, the detailed balance for such a process must be satisfied at any point in time.
In all the TM scenario considered in this work, the `spectral diffusion' mechanism depicted here below
\begin{equation}
\begin{array}{c c c c}
\downarrow & \uparrow & \uparrow & \downarrow	\\
\omega_\text{i-$\delta$} & \omega_\text{i} & \omega_\text{j} & \omega_\text{j+$\delta$}
\end{array} 
\begin{array}{c}
	\;\;\rightleftharpoons \;\;\\
	\;\;	T_{2e}\;\;
\end{array}
\begin{array}{c c c c}
\uparrow & \downarrow & \downarrow & \uparrow	\\
\omega_\text{i-$\delta$} & \omega_\text{i} & \omega_\text{j} & \omega_\text{j+$\delta$}
\end{array}  \nonumber
\label{MW}
\end{equation}		
has been always assumed to be a fast process. Its corresponding detailed balance condition can be written in terms of the fraction of electrons \emph{up} - $P_{e, i}^+(t)$ - and of the fraction of electrons \emph{down} - $P_{e, i}^-(t)$ - at time $t$ : 
\begin{equation}
\frac{P_{e, i\small{-}\delta}^+(t)P_{e, i}^-(t)}{P_{e, i\small{-}\delta}^-(t)P_{e, i}^+(t)}=\frac{P_{e, j}^+(t)P_{e, j\small{+}\delta}^-(t)}{P_{e, j}^-(t)P_{e, j\small{+}\delta}^+(t)}. \nonumber
\end{equation}
Then, by using the relation 
\begin{equation}
P_{e, i}^+(t)=\frac{1+P_{e, i}(t)}{2}, \quad P_{e, i}^-(t)=\frac{1-P_{e, i}(t)}{2}   \nonumber
\end{equation}
one comes to an equation for the electron polarization
\begin{equation}
\frac{(1-P_{e,i})(1+P_{e, i-\delta})}{(1+P_{e,i})(1-P_{e, i-\delta})}= \frac{(1+P_{e,j})(1-P_{e, j+\delta})}{(1-P_{e,j})(1+P_{e, j+\delta})}\nonumber
\end{equation} 
which is satisfied if 
\begin{equation}
P_{e, i}(t) = \tanh[\beta_e(t)(\Delta_i -c(t))],\nonumber
\end{equation}
$qed$.

Repeating the same procedure for the ISS process one obtains the following equation for the nuclear polarization:
\begin{equation} \label{A2}
P_n(t) = \frac{P_{e, i}(t)-P_{e,i+\delta n_p}(t)}{1-P_{e,i+\delta n_p}(t) \, P_{e, i}(t)}
\end{equation}
which, being $P_{e, i}(t)=\tanh\left[\beta(t)(\Delta_i-c(t))\right]$, can be rewritten as:
\begin{equation} \label{A3}
P_n(t) = \tanh\left[ \beta(t) \delta n_p \right]. 
\end{equation}
Eq.(\ref{A3}), valid when $T_{\text{ISS}}$ is as fast as spectral diffusion, defines the existence of a unique common temperature between nuclear spin system and electron non Zeeman reservoir.

\section{Electron and nuclear spin dynamics}
\label{appDynamics}
The numerical procedure described in the main text to evaluate $P_{e, i}(t)$ and $P_n(t)$, when all the processes but the spectral diffusion have a finite transition rate, has been conveniently adapted for the three regimes considered. 
The strategy consists in using conservation laws to manage all mechanisms assumed to be  infinitely efficient, while computing rate equations only for processes with finite rate.

\subsection{Regime I: `Borghini' ($T_{\text{ISS}}$ and $T_{1\text{MW}} \rightarrow 0$)}
\label{appDynamicsI}
Rate equations are used to account only for the effect of the electron and nuclear spin-lattice relaxation:
\begin{eqnarray}
\label{rateeqBorghini}
\frac{dP_{e, i}(t)}{dt} = \frac{P_0-P_{e, i}(t)}{T_{1e}} \nonumber \\
\frac{dP_n(t)}{dt}= \frac{P_{0n}-P_n(t)}{T_{1n}}  
\end{eqnarray}
Fast processes (spectral diffusion, electron-nucleus contact and MW saturation) are accounted by the conservation of the total polarization (for variations induced by spectral diffusion or ISS) and of the total electron non Zeeman plus nuclear Zeeman energies: 
\begin{eqnarray}
&&\sum f_i \left[P_{e, i}(t + \delta t)-P_{e, i}(t)-\delta_{i, i_0}\delta P^{MW}\right] = 0 \nonumber \\
&&\sum f_i \Delta_i \left[P_{e, i}(t\small{+}\delta t)\small{-}P_{e, i}(t)\right]\small{-}\frac{N_n}{N_e}\omega_n\left[P_n (t\small{+}\delta t) \small{-}P_n (t) \right] \small{=} 0 \nonumber
\end{eqnarray}
where $\delta P^{MW}$ indicates the variation due to MW irradiation and the time step $\delta t \rightarrow 0$ being the characteristic time of the transitions $T_{2e}$, $T_{\text{ISS}}$ and $T_{1\text{MW}}$.
These equations are conveniently written as:
\begin{eqnarray}
&&\small{-}\sum_{i\neq i_0} f_i \left[P_{e, i}(t\small{+}\delta t)\small{-}P_{e, i}(t)\right] = f_0 \left[P_{e, 0}(t\small{+}\delta t)\small{-}P_{e, 0}(t)\right.\nonumber \\
&&\left.-\delta P^{MW}\right] \nonumber \\
&&\sum_{i \neq i_0} f_i \Delta_i \left[P_{e, i}(t\small{+}\delta t)\small{-}P_{e, i}(t)\right]\small{+}f_0 \Delta_0 \left[P_{e, 0}(t\small{+}\delta t)\small{-}P_{e, 0}(t)\right.\nonumber \\ 
&&\left.-\delta P^{MW}\right]\small{-} \frac{N_n}{N_e} \omega_n \left[P_n (t\small{+}\delta t) \small{-} P_n (t)\right] = 0\nonumber 
\end{eqnarray}
so that, by means of simple algebra, the condition:
\begin{eqnarray}  
\label{CLBorghini}
&&\sum f_i (\Delta_i\small{-}\Delta_0) \left[P_{e, i}(t \small{+} \delta t)\small{-}P_{e, i}(t)\right]\nonumber \\
&\small{-}& \frac{N_n}{N_e} \omega_n \left[P_n (t \small{+}\delta t) \small{-} P_n (t) \right]= 0
\end{eqnarray}
is obtained.
By solving this equation one derives $\beta(t+\delta t)$ and computes $P_{e, i}(t+\delta t) = -\tanh\left[\beta(t+\delta t)(\Delta_i -\Delta_0)\right]$ and $P_n(t+\delta t) = \tanh\left[\beta(t+\delta t)\delta n_p\right]$.

It is interesting to study also the evolution of the inverse temperature $\beta(t)$ that in regime I, as demonstrated in the Appendix \ref{appTM}, is the same for both the electron non Zeeman and the nuclear Zeeman reservoirs: $\beta(t)=\beta_e(t)=\beta_n(t)$. Moreover, since full saturation imposes $c = \Delta_0$, $\beta(t)$ is the only unknown variable of the problem. Hence, by means of Eq.(\ref{CLBorghini}) and Eq.(\ref{rateeqBorghini}), it is possible to describe analitically the time behaviour of $\beta(t)$.
At a generic time $t+ dt$, by assuming $\delta t \rightarrow 0$, Eq.(\ref{CLBorghini}) writes:
\begin{eqnarray}
&& \sum f_i \left\{ -\tanh\left[ \beta(t+dt) \left( \Delta_i -\Delta_0\right)\right] -P_{e, i}(t+dt)\right\} \nonumber \\
&& \left( \Delta_i\small{-}\Delta_0\right)\small{-}\frac{N_n}{N_e} \omega_n \left[\tanh\left(\beta(t+dt) \omega_n\right)-P_n(t+dt)\right] = 0 \nonumber
\end{eqnarray}
Now, using Eq.(\ref{rateeqBorghini}) for replacing $P_{e, i}(t+dt)$ and $P_n(t+dt)$ one obtains:
\begin{eqnarray}
\label{betaBorghini3}  
&& \sum f_i \left\{ -\tanh \left[ \beta(t+dt) \left( \Delta_i -\Delta_0\right)\right]+ \right. \nonumber \\
&&+ \tanh \left[ \beta(t) \left( \Delta_i -\Delta_0\right)\right] +  \nonumber \\
&& \left. - \frac{dt}{T_{1e}} P_0 - \tanh \left[ \beta(t) \left( \Delta_i -\Delta_0\right)\right] \right\}\left( \Delta_i -\Delta_0\right)+ \nonumber \\
&& - \frac{N_n}{N_e} \omega_n \left[\tanh\left(\beta(t+dt) \omega_n\right)-\tanh\left(\beta(t) \omega_n\right) \right] \nonumber \\
&& = 0 \nonumber
\end{eqnarray}
Then, with the first order expansions:
\begin{eqnarray}
&&\beta(t\small{+}dt) \approx \beta(t)\small{+}\beta'(t)dt \nonumber \\
&&\tanh\left[\beta(t\small{+}dt)x\right]\small{\approx}\tanh\left[\beta(t)x\right]\small{+}\beta'(t) x \left\{1\small{-}\tanh^2\left[\beta(t)x\right]\right\} \nonumber
\end{eqnarray}
and some algebric calculations, the following equation for $\beta(t)$ is achieved:
\begin{eqnarray}
\label{betaeq}  
&& \beta'(t) \left\{ \sum f_i \left(\Delta_i -\Delta_0 \right)^2 \left[1\small{-}\tanh^2\left[\beta(t)\left(\Delta_i -\Delta_0 \right)\right]\right] + \right. \nonumber \\
&& \left. \frac{N_n}{N_e}\omega_n^2 \left[1\small{-}\tanh^2\left[\beta(t)\omega_n\right]\right]\right\} +  \\
&& \frac{1}{T_{1e}} \sum f_i \left(\Delta_i -\Delta_0 \right) \left\{P_0 + \tanh \left[ \beta(t) \left( \Delta_i -\Delta_0 \right)\right] \right\} = 0 \nonumber
\end{eqnarray}
Eq.(\ref{betaeq}) is conveniently rewritten as:
\begin{equation}
\label{betaeqfinal}
\beta'(t) = - \frac{\sum f_i \left(\Delta_i -\Delta_0 \right)  \tanh \left[ \beta(t) \left( \Delta_i -\Delta_0 \right)\right] -\Delta_0 P_0}{\frac{N_n}{N_e T_{1e}}\omega_n^2\left[1\small{-}\tanh^2\left[\beta(t)\omega_n\right]\right]},
\end{equation}
after negleting with good approximation the term $\sum f_i\left(\Delta_i\small{-}\Delta_0\right)^2\left[1\small{-}\tanh^2\left[\beta(t)\left(\Delta_i\small{-}\Delta_0 \right)\right]\right]$.

Although no attemp to solve analically Eq.(\ref{betaeqfinal}) is done here, it is clear that the solution can not be an exponential function, as already anticipated in the main text.

\subsection{Regime II: partial MW saturation ($T_{2e}$, $T_{\text{ISS}} \rightarrow 0$)}
\label{appDynamicsII}
The system of rate equations is used to describe the effect of partial MW saturation as well as electron and nuclear spin-lattice relaxation:
\begin{eqnarray}
\label{rateeqII}
\frac{dP_{e, i}(t)}{dt} &=& \frac{P_0-P_{e, i}(t)}{T_{1e}} - \delta_{i, i_0} \frac{P_{e, 0}}{T_{1\text{MW}}} \nonumber \\
\frac{dP_n(t)}{dt} &=& \frac{P_{0n}-P_n(t)}{T_{1n}}  \nonumber
\end{eqnarray}
whereas spectral diffusion and electron-nucleus interaction are accounted by the following conservation laws:
\begin{eqnarray}
&&\sum f_i \left[P_{e, i}(t + \delta t)-P_{e, i}(t)\right] = 0 \nonumber \\
&&\sum f_i \Delta_i \left[P_{e, i}(t\small{+}\delta t)\small{-}P_{e, i}(t)\right]\nonumber\\
&&\small{-} \frac{N_n}{N_e}\omega_n\left[P_n (t\small{+}\delta t) \small{-}P_n (t) \right] = 0 \nonumber
\end{eqnarray}
with $\delta t \rightarrow 0$ being the characteristic time of the transitions $T_{2e}$ and $T_{\text{ISS}}$.
By solving this system one obtains $\beta(t+\delta t)$ and $c(t+\delta t)$ and computes $P_{e, i}(t+\delta t) = -\tanh\left[\beta(t+\delta t)(\Delta_i -c(t))\right]$ and $P_n(t+\delta t) = \tanh\left[\beta(t+\delta t)\delta n_p\right]$.

\subsection{Regime III: poor electron-nucleus contact ($T_{2e}$ and $T_{1\text{MW}} \rightarrow 0$)}
\label{appDynamicsIII}
The system of rate equations takes into account the effect of the electron-nucleus contact and of the electron and nuclear spin-lattice relaxation:
\begin{eqnarray}
 \frac{d P_{e, i}(t)}{dt}&=& \frac{P_0-P_{e, i}(t)}{T_{1e}} \nonumber \\
 &+& \frac{f_{i-\delta n_p} \Pi_- + f_{i+\delta n_p} \Pi_+ }{2 T_{\text{ISS}}}  \nonumber \\
 \frac{d P_n(t)}{dt}&=& \frac{P_{0n}-P_n(t)}{T_{1n}}- \frac{N_e}{2 T_{\text{ISS}} N_n} \sum f_i f_{i+\delta n_p} \Pi_n  \nonumber
\end{eqnarray}
The effect of the other processes (spectral diffusion and full MW saturation) is accounted by the conservation of the total polarization (when the variation is induced by spectral diffusion) and of the total electron non Zeeman plus nuclear Zeeman energies:
\begin{eqnarray}
&&\sum f_i \left[P_{e, i}(t + \delta t)-P_{e, i}(t)-\delta_{i,i_0}\delta P^{MW}\right] = 0 \nonumber \\
&&\sum f_i \Delta_i \left[P_{e, i}(t\small{+}\delta t)\small{-}P_{e, i}(t)\right] = 0 \nonumber
\end{eqnarray}
where $\delta P^{MW}$ indicates the variation due to MW irradiation and the time step $\delta t \rightarrow 0$ being the characteristic time of the transitions $T_{2e}$ and $T_{1\text{MW}}$.
These equations are conveniently written as:
\begin{eqnarray}
&&-\sum_{i\neq i_0} f_i \left[P_{e, i}(t\small{+}\delta t)\small{-}P_{e, i}(t)\right] = f_0 \left[P_{e, 0}(t\small{+}\delta t)\small{-}P_{e, 0}(t)\right. \nonumber \\
&&\left.-\delta P^{MW}\right] \nonumber \\
&&\sum_{i \neq i_0} f_i \Delta_i \left[P_{e, i}(t\small{+}\delta t)\small{-}P_{e, i}(t)\right]\small{+}f_0 \Delta_0 \left[P_{e, 0}(t\small{+}\delta t)\small{-}P_{e, i}(0)\right. \nonumber \\
&&\left.-\delta P^{MW}\right] = 0\nonumber 
\end{eqnarray}
so that, after simple algebric calculations, the following condition is obtained:
\begin{eqnarray}  
&&\sum f_i (\Delta_i\small{-}\Delta_0) \left[P_{e, i}(t \small{+} \delta t)\small{-}P_{e, i}(t)\right] = 0, \nonumber
\end{eqnarray}
that allows deriving $\beta(t+\delta t)$ and thus computing $P_{e, i}(t+\delta t) = -\tanh\left[\beta(t+\delta t)(\Delta_i -\Delta_0)\right]$.

\section{Dynamical behaviour of the electron average polarization}
\label{kineticPe}
\begin{figure*}[htbp]
 \includegraphics[width=15.6 cm]{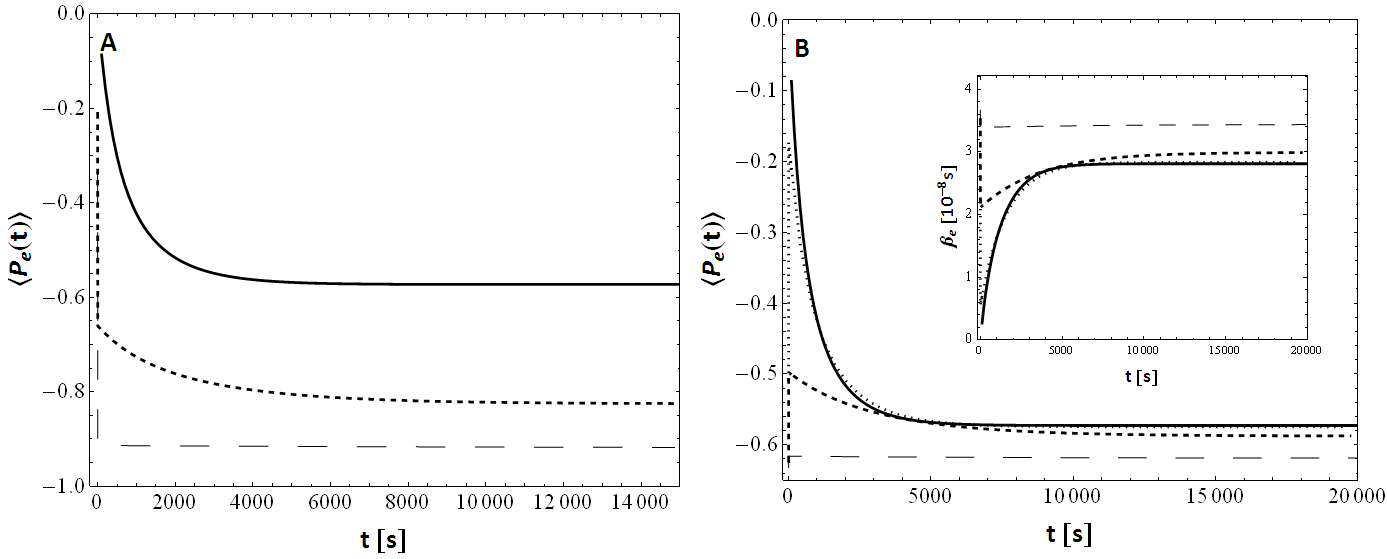}
\caption{Panel A: Build up curve of the average electron polarization $\langle P_{e}(t) \rangle$ in regime of partial saturation (regime II) for $T_{1\text{MW}} = 0$ s (Borghini limit, thick solid line), 0.1 s (small-dashed line) and 1 s (large-dashed line). Remaining parameters are set as follows: $T_{1e}$ = 1 s, $T_{1n} = \infty$, $N_n/N_e$ = 1000, $i_0$ =4, $\delta n_p$ = 3, $N_p$ = 15. Panel B: Build up curve of the average electron polarization $\langle P_{e}(t) \rangle$ and of the inverse electron spin temperature $\beta_e(t)$ (inset) in regime of poor electron-nucleus contact for $T_{\text{ISS}} = 0$ s (Borghini limit, thick solid line), 0.01 s (dotted line), 0.1 s (small-dashed line) and 1 s (large-dashed line). Remaining parameters are set as follows: $T_{1e}$ = 1 s, $T_{1n}$ = 10000 s, $N_n/N_e$ = 1000, $i_0$ = 5, $\delta n_p$ = 3, $N_p$ = 15.}
\label{figurePeDyn}
\end{figure*}

The evolution of the average electron polarization $\langle P_e \rangle$ in regime II and regime III shows a peculiar behaviour characterized by two different time scales, as sketched in \figurename~\ref{figurePeDyn}. 
\subsection{Partial MW saturation}
In this regime, due to the hypothesis $T_{2e} = 0$ and $T_{\text{ISS}} = 0$, one has $P_{e,i}(t) = \tanh\left[\beta_e(t)\left(\Delta_i- c(t)\right)\right]$ and $\beta_e(t) = \beta_n(t) = \beta(t)$. The evolution of $\langle P_e \rangle$ is determined both by $\beta(t)$ (panel D, \figurename~\ref{figureII}) and $c(t)$. 

At short times ($t \approx T_{1e}$) the inverse temperature $\beta$ is determined by the large nuclear system for which $\beta_n(t=0)=\beta_L \approx 0$. 
When $T_{1\text{MW}} = 0$, the only solution is $c(t=0)=\Delta_0$ and consequently $P_{e, i}(t=0) = 0, \forall i$. For partial saturation ($T_{1\text{MW}}>0$) the profile of $P_{e, i}$ becomes a flat function (corresponding to the condition $c(t=0) \rightarrow \infty$) which quickly evolves with a characteristic time $T_{1e}$ towards an intermediate level between 0 and $P_0$, that can be calculated using the first equation of system \ref{BorghiniRelationMW}:
\begin{equation}
P_{e, i} = P_0 \frac{T_{1\text{MW}}}{f_0 T_{1e}+T_{1\text{MW}}}. \nonumber
\end{equation}

At longer times ($t \approx T_{\text{pol}}$) the evolution of $\langle P_e \rangle$ is mainly due to $\beta(t)$ dynamics  (being $c(t)$ approximately constant) and it is thus characterized by a time constant in the order of $10^3$ s.

\subsection{Poor electron-nucleus contact}
The dynamics of both $\beta_e(t)$ and $\langle P_e(t) \rangle$ is characterized by two time scales: a first rapid component with a characteristic time in the order of $T_{1e}$ and a second slow component with a characteristic time in the order of $T_{\text{pol}}$. 

In this regime, due to the hypothesis $T_{2e} = 0$ and $T_{1\text{MW}} = 0$, one has $P_{e,i}(t) = \tanh \left[\beta_e(t) (\Delta_i- \Delta_0)\right]$, with $\beta_e(t) \neq \beta_n(t)$. Depending on the time scale considered, the system can be qualitatively depicted and $\beta_e$ estimated accordingly. 
\begin{itemize}
\item At very short times ($t \rightarrow 0$), being the contact between electrons and nuclei finite, the electron system is unaffected by the presence of the nuclear reservoir and reaches immediately the inverse temperature $\beta_B$ predicted by Borghini and defined by Eq.(\ref{CLBorghini}) after setting $N_n = 0$.
\item After this initial `thermalization' phase, at times $t \approx T_{1e}$, the electron reservoir is on one side in contact with a thermal bath at temperature $1/\beta_B$ (determined by interaction with the lattice, by spectral diffusion and by the highly effective MWs), while feeling on the other side the nuclear ensamble having an initial temperature $\beta_n = \beta_L \approx 0$. Thus, on a time scale of few $T_{1e}$, the inverse temperature $\beta_e$ moves towards a target value between $\beta_B$ and $\beta_L$, depending on the strenght of the two contact times $T_{1e}$ and $T_{\text{ISS}}$. When the electron-nucleus contact is poorly efficient $\beta_e \rightarrow \beta_B$; conversely $\beta_e \rightarrow \beta_L \approx 0$  for strong electron-nucleus contact. 
\item At large times ($t \approx T_{\text{pol}}$), $\beta_n(t)$ evolves from $\beta_L$ towards its final steady state $\beta_n$ and $\beta_e(t)$ evolves as well, reaching an intermediate value between $\beta_n$ and $\beta_B$.
\end{itemize}
\vspace{0.03cm}
In summary, as long as the electron-nucleus contact is poorly efficient, the electron inverse temperature $\beta_e$ is only slightly affected by the nuclear reservoir and it is thus seen by this latter as a constant value equal to $\beta_B$. 
As discussed in Section \ref{sectionModel} and in Section \ref{Discussion} this behaviour leads streightforward to an exponential build up curve for nuclear polarization.


\begin{thebibliography}{0}
\bibitem{PNAS JHAL} J. H. Ardenkjaer-Larsen, B. Fridlund, A. Gram, G. Hansson, L. Hansson, M. H. Lerche, R. Servin, M. Thaning and K. Golman, Proc. Natl. Acad. Sci. {\bf100}, 10158 (2003).
\bibitem{SS1} D. Hall, D. Maus, G. Gerfen, S. Inati, L. Becerra, F. Dahlquist, R. Griffin, Science {\bf276} 930 (1997).
\bibitem{SS2} M. Rosay, J. Lansing, K. Haddad, W. Bachovchin, J. Herzfeld, R. Temkin, R. Griffin, J. Am. Chem. Soc. {\bf125} 13626 (2003).
\bibitem{BA1} K. Golman, R. in�t Zandt, M. Lerche, R. Pehrson and J. H. Ardenkjaer-Larsen, Cancer Res {\bf22} 66 (2006).
\bibitem{BA2} S. E. Day, M. I. Kettunen, F. A. Gallagher, D. Hu, M. Lerche, J. Wolber, K. Golman, J. H. Ardenkjaer-Larsen and K. M. Brindle, Nat. Med. {\bf13} 1382 (2007).
\bibitem{BA3} J. Kurhanewicz, D, B. Vigneron, K. Brindle, E, Y. Chekmenev, A. Comment, C. H. Cunningham, R. J. DeBerardinis, G. G. Green, M. O. Leach, S. S. Rajan, R. R. Rizi, B. D. Ross, W. S.Warren and C. R. Malloy, Neoplasia {\bf13} 81 (2011).
\bibitem{Vega1} Y. Hovav, A. Feintuch and S. Vega, J. Chem. Phys. {\bf134}, 074509 (2011).
\bibitem{kock} A. Karabanov, A. van der Drift, L. J. Edwards, I. Kuprovb and W. K$\ddot{o}$ckenberger, Phys. Chem. Chem. Phys. {\bf14}, 2658 (2012).
\bibitem{Vega3} Y. Hovav, A. Feintuch and S. Vega, J. Magn. Reson. {\bf214}, 29 (2012). 
\bibitem{khut} J.R. Khutsishvili Soviet Physics Uspekhi, {\bf8}, 747 (1966).
\bibitem{Abragam e Goldman} A. Abragam and M. Goldman, Nuclear magnetism: order and disorder. Oxford: Clarendon Press, (1982).
\bibitem{CE1} A. V. Kessenikh, V. I. Lushchikov, A. A. Manekov and Y. V. Taran, Sov. Phys. {\bf5}, 321 (1963).
\bibitem{CE2} A. V. Kessenikh, A. A. Manekov and G. I. Pyatnitskii, Sov. Phys. {\bf6}, 641 (1964).
\bibitem{CE3} C. F. Hwang and D. A. Hill, Phys. Rev. Lett. {\bf18}, 110 (1967).
\bibitem{CE4} C. F. Hwang and D. A. Hill, Phys. Rev. Lett. {\bf19}, 1011 (1967).
\bibitem{Borghini PRL} M. Borghini, Phys. Rev. Lett. {\bf20}, 419 (1968).  
\bibitem{Weck} W.T. Wenckebach, T.J.B. Swaneburg and N.J. Poulis, Physics reports {\bf14} 181 (1974).
\bibitem{JHAL2008} J.H. Ardenkjaer-Larsen, S. Macholl and H. Johannesson, App. Magn. Reson. {\bf34}, 509 (2008).
\bibitem{JanninMW} S. Jannin, A. Comment and J. J. van der Klink, Appl. Mag. Res. {\bf43}, 59 (2012). 
\bibitem{nostroPCCP} S. Colombo Serra, A. Rosso and F. Tedoldi, Phys. Chem. Chem. Phys. {\bf14}, 13299 (2012).
\bibitem{JHAL2010} S. Macholl, H. Johannesson and J.H. Ardenkjaer-Larsen, Phys. Chem. Chem. Phys. {\bf12}, 5804 (2010).
\bibitem{JHALhighfield} H. Johannesson, S. Macholl and J. H. Ardenkjaer-Larsen, J. Magn. Reson. {\bf197}, 167 (2009).
\bibitem{Jannin2008} S. Jannin, A. Comment, F. Kurdzesau, J. A. Konter, P. Haute, B. van den Brandt and J. J. van der Klink, The Journal of Chemical Physics {\bf128}, 241102 (2008).
\bibitem{Griffin1} L. Becerra, G. Gerfen, R. Temkin, D. Singel, R. Griffin, Phys Rev Lett. {\bf71}, 3561 (1993).
\bibitem{Griffin2} V. Bajaj, C. Farrar, M. Hornstein, I. Mastovsky, J. Vieregg, J. Bryant, B. Elena, K. Kreischer, R. Temkin, R. Griffin, J Magn Reson. {\bf160}, 85 (2003).
\bibitem{Griffin3} A. B. Barnes, E. Markhasin, E. Daviso, v. K. Michaelis, E. A. Nanni, S. K. Jawla, E. L. Mena, R. DeRocher, A. Thakkar, P. P. Woskov, J. Herzfeld, R. Temkin., R. Griffin, J Magn Reson. {\bf224}, 1 (2012).
\end{thebibliography}
\end{document}